\documentclass{article}
\usepackage{amssymb}
\usepackage{amsfonts}
\usepackage{amsmath}
\usepackage[doublespacing]{setspace}

\input{tcilatex}

\begin{document}

\title{A quasi-free position-dependent-mass jump and self-scattering
correspondence}
\author{Omar Mustafa$^{1}$ and S.Habib Mazharimousavi$^{2}$ \\
Department of Physics, Eastern Mediterranean University, \\
G Magusa, North Cyprus, Mersin 10,Turkey\\
$^{1}$E-mail: omar.mustafa@emu.edu.tr, $^{2}$E-mail: habib.mazhari@emu.edu.tr%
\\
\ Tel: +90 392 630 1314, Fax: +90 392 3651604}
\maketitle

\begin{abstract}
A quasi-free quantum particle endowed with Heaviside position dependent mass
jump is observed to experience scattering effects manifested by its
by-product introduction of the derivative of the Dirac's-delta point dipole
interaction, $\delta ^{\prime }\left( x\right) =\partial _{x}\delta \left(
x\right) $. Using proper parametric mappings, the reflection and
transmission coefficients are obtained. A new ordering ambiguity parameters
set, as the only feasibly admissible within the current methodical proposal,
is suggested.


\end{abstract}

\section{Introduction}

Hamiltonians for particles endowed with position-dependent-mass (PDM) (i.e., 
$M\left( x\right) =m_{\circ }m\left( x\right) $) have attracted much
research attention over the last few decades [1-29]. Such attention was
inspired not only by the feasible applicability of PDM-settings in the study
of various physical problems (e.g., many-body problem, semiconductors,
quantum dots, quantum liquids, etc.) but also by the mathematical challenge
associated with the ordering ambiguity in the PDM van Roos Hamiltonian. The
non-commutativity between the momentum operator (with $\hbar =m_{\circ }=1$
units to be used through out) $\hat{p}_{x}=-i\partial _{x}$ and the
position-dependent-mass results in an ordering ambiguity in the
non-uniqueness representation of the kinetic energy operator%
\begin{equation*}
T=-\frac{1}{4}\left[ M\left( x\right) ^{\alpha }\partial _{x}M\left(
x\right) ^{\beta }\partial _{x}M\left( x\right) ^{\gamma }+M\left( x\right)
^{\gamma }\partial _{x}M\left( x\right) ^{\beta }\partial _{x}M\left(
x\right) ^{\alpha }\right] ,
\end{equation*}%
where $\alpha $, $\beta $, and $\gamma $ are called the van Roos ordering
ambiguity parameters satisfying the van Roos constraint $\alpha +\beta
+\gamma =-1$ [cf., e.g., 25-29].

In the literature, there exist several suggestions for the van Roos ordering
ambiguity parameters. Amongst, the Gora's and Williams' ($\beta =\gamma =0,$ 
$\alpha =-1$), Ben Daniel's and Duke's ($\alpha =\gamma =0,$ $\beta =-1$),
Zhu's and Kroemer's ($\alpha =\gamma =-1/2,$ $\beta =0$) , Li's and Kuhn's ($%
\beta =\gamma =-1/2,$ $\alpha =0$), and the very recent Mustafa's and
Mazharimousavi's ($\alpha =\gamma =-1/4,$ $\beta =-1/2$) (cf. e. g., [10,
29] for more details on this issue). It has been observed (cf., e. g.,
[29,40]) that the physical and/or mathematical admissibility of a given
ambiguity parameters set very well depends not only on the continuity
conditions at the abrupt heterojunction boundaries but also on the
position-dependent-mass form and/or potential form. The general consensus is
that there is no unique neither there is a universal choice for these
ambiguity parameters, therefore.

On the other hand, research activities on the analysis of the
one-dimensional Hamiltonians associated with the what is called "point" or
"contact" interactions (i.e., zero everywhere except at the origin $x=0$,
like Dirac delta $\delta \left( x\right) $ distribution), in solid-state
physics, were stimulated by the rapid progress in the fabrication of
nanoscale quantum devices [30-36]]. Such interactions are intuitively
understood as sharply localized potentials exhibiting a number of
interesting features. Their feasible applicability extends to optics when
electromagnetic waves scatter at the boundaries of thin layers in dielectric
media [37].

Within the context of the recent interest in exactly solvable
one-dimensional Schr\"{o}dinger models of scattering accompanied by
position-dependent-mass particles, we consider, in this letter, a quasi-
free particle (i.e., subjected to $V\left( x\right) =0$ potential)\ endowed
with a Heaviside step mass function/distribution of the form 
\begin{equation}
m\left( x\right) =1+\mu \emph{h}\left( x\right) \text{ };\text{ }%
\mathbb{R}
\ni \mu >0.
\end{equation}%
Here%
\begin{equation*}
\emph{h}\left( x\right) =\frac{1+sgn\left( x\right) }{2}=\left\{ 
\begin{tabular}{ll}
$0$ & ; $x<0$ \\ 
$1/2$ & ; $x=0$ \\ 
$1$ & ; $x>0$%
\end{tabular}%
\right.
\end{equation*}%
is the discontinuous Heaviside step function. Nevertheless, for the
convenience of the current study we shall use a more general form for the
PDM-function to read%
\begin{equation}
m\left( x\right) =f\left( \emph{h}\left( x\right) \right) =\left\{ 
\begin{tabular}{ll}
$m_{1}$ & ; $x<0$ \\ 
$m_{2}$ & ; $x=0$ \\ 
$m_{3}$ & ; $x>0$%
\end{tabular}%
\right. .
\end{equation}%
Of course this would practically refer to "position-dependent-mass jumps"
(cf., e.g., [38-40]). To the best of our knowledge, such unusual PDM
settings of a Heaviside discontinuous functional nature have been discussed
in the literature (cf., e.g., [38-40]), but never within our forthcoming
methodical proposal, at least. It would be interesting to subject such mass
settings to the sequel theoretical experiment, therefore.

We witness (in section 2) that a quasi-free quantum particle (i.e.,
subjected to $V\left( x\right) =0$ whilst $V_{eff}\left( q\left( x\right)
\right) \neq 0$) endowed with the PDM-setting of (2) would experience
scattering effects manifested by the particle's by-product introduction of
the derivative of the Dirac's delta interaction $\delta ^{\prime }\left(
x\right) $ as a result of a point canonical transformation (PCT) recipe
(hence, the notion of self-scattering correspondence is unavoidable). The
detailed solution of which can be inferred from the scattering potential $%
V\left( q\right) =-a\delta \left( q\right) +b\delta ^{\prime }\left(
q\right) $ of Gadella et al. [36] using proper parametric mappings into our
model (see (14) below), of course. In this case, the reader may wish to
refer to Gadella et al. [36] for the mathematical and/or physical details.
Moreover, a new (the only feasibly admissible within the current methodical
proposal) ordering ambiguity parameters set obtains in the process. We
conclude in section 3.

\section{PCT recipe and self-scattering correspondence}

Under position-dependent-mass settings, the von Roos PDM Schr\"{o}dinger
equation [22-29] (in $\hslash =m_{\circ }=1$ units) reads%
\begin{equation}
\left[ -\frac{1}{2}\partial _{x}\left( \frac{1}{m\left( x\right) }\right)
\partial _{x}+\tilde{V}\left( x\right) \right] \psi \left( x\right) =E\psi
\left( x\right) ,
\end{equation}%
with%
\begin{equation}
\tilde{V}\left( x\right) =g_{1}\frac{m^{\prime \prime }\left( x\right) }{%
m\left( x\right) ^{2}}-g_{2}\frac{m^{\prime }\left( x\right) ^{2}}{m\left(
x\right) ^{3}}
\end{equation}%
where primes denote derivatives with respect to $x$ and%
\begin{equation}
g_{1}=\frac{1}{4}\left( 1+\beta \right) \text{ ; \ }g_{2}=\frac{1}{2}\left[
\alpha \left( \alpha +\beta +1\right) +\beta +1\right] .
\end{equation}%
We now follow the well-known point-canonical-transformation (PCT) recipe
(cf. e.g., [19]) through the substitution $\psi \left( x\right) =m\left(
x\right) ^{1/4}\phi \left( q\left( x\right) \right) $ in (3) to imply (with $%
q^{\prime }\left( x\right) =\sqrt{m\left( x\right) }$)%
\begin{equation}
q\left( x\right) =\int^{x}\sqrt{m\left( t\right) }dt=\int^{x}\sqrt{f\left( 
\emph{h}\left( t\right) \right) }dt=x\,\sqrt{f\left( \emph{h}\left( x\right)
\right) },
\end{equation}%
and obtain a Schr\"{o}dinger equation of the form%
\begin{equation}
\left[ -\frac{1}{2}\partial _{q}^{2}+V_{eff}\left( q\right) \right] \phi
\left( q\right) =E\phi \left( q\right) ,
\end{equation}%
where%
\begin{equation}
V_{eff}\left( q\right) =\tilde{V}\left( x\right) +\frac{1}{2}\left( \frac{%
7m^{\prime }\left( x\right) ^{2}}{32m\left( x\right) ^{3}}-\frac{m^{\prime
\prime }\left( x\right) }{8m\left( x\right) ^{2}}\right) .
\end{equation}%
Which would in turn (with $m^{\prime }\left( x\right) =\partial _{x}m\left(
x\right) ,$ and $\,f^{\prime }\left( \emph{h}\left( x\right) \right)
=\partial _{\emph{h}\left( x\right) }f\left( \emph{h}\left( x\right) \right) 
$ ) yield 
\begin{align}
V_{eff}\left( q\left( x\right) \right) & =G_{1}\frac{m^{\prime \prime
}\left( x\right) }{m\left( x\right) ^{2}}-G_{2}\frac{m^{\prime }\left(
x\right) ^{2}}{m\left( x\right) ^{3}}  \notag \\
& =G_{1}\frac{\delta ^{\prime }\left( x\right) \,f^{\prime }\left( \emph{h}%
\left( x\right) \right) }{\,f\left( \emph{h}\left( x\right) \right) ^{2}}+%
\frac{\delta \left( x\right) ^{2}\,}{\,f\left( \emph{h}\left( x\right)
\right) ^{2}}\left[ G_{1}f^{\prime \prime }\left( \emph{h}\left( x\right)
\right) -G_{2}\frac{\,f^{\prime }\left( \emph{h}\left( x\right) \right) ^{2}%
}{\,f\left( \emph{h}\left( x\right) \right) }\right] ,
\end{align}%
with%
\begin{equation}
G_{1}=\frac{1}{8}\left( 1+2\beta \right) \text{ ; \ }G_{2}=\frac{1}{2}\left[
\alpha \left( \alpha +\beta +1\right) +\beta +\frac{9}{16}\right]
\end{equation}%
Where $\delta \left( x\right) $ denotes Dirac delta function and $\delta
^{\prime }\left( x\right) =\partial _{x}\delta \left( x\right) $ is the
derivative of the Dirac delta function. Hereby, the terms associated with $%
\delta ^{\prime }\left( x\right) $ and $\delta \left( x\right) ^{2}$ form
continuous functions except at the origin $x=0$ and are of a finite
discontinuity, therefore. Under these settings, we may benefit from the
well-known definitions associated with the Dirac delta distributions (cf.,
e.g., equations (4) and (5) of [36]). That is, if $U\left( x\right) $ is a
discontinuous function of $x$ then the distributions $U\left( x\right)
\delta \left( x\right) $ and $U\left( x\right) \delta ^{\prime }\left(
x\right) $ can be rewritten as 
\begin{equation}
U\left( x\right) \delta \left( x\right) =U\left( 0\right) \delta \left(
x\right) ,
\end{equation}%
\begin{equation}
U\left( x\right) \delta ^{\prime }\left( x\right) =U\left( 0\right) \delta
^{\prime }\left( x\right) -U^{\prime }\left( 0\right) \delta \left( x\right)
.
\end{equation}%
Which would imply (with $U\left( x\right) =f^{\prime }\left( \emph{h}\left(
x\right) \right) /\,f\left( \emph{h}\left( x\right) \right) ^{2}$) that the
effective potential in (9) can be recast as 
\begin{equation}
V_{eff}\left( q\left( x\right) \right) =G_{1}U\left( 0\right) \delta
^{\prime }\left( x\right) +\left( 2G_{1}-G_{2}\right) \frac{\,f^{\prime
}\left( \emph{h}\left( x\right) \right) ^{2}}{\,f\left( \emph{h}\left(
x\right) \right) ^{3}}\delta \left( x\right) ^{2}.
\end{equation}

To avoid the physical and/or mathematical meaninglessness of $\delta \left(
x\right) ^{2}$, two feasible solutions for (13) obtain. The simplest of
which is achieved by taking $G_{1}=0$ and $G_{2}=0$ (i.e., Mustafa and
Mazharimousavi's [10], MM-, ordering ambiguity parameters $\alpha =\gamma
=-1/4$ and $\beta =-1/2$ here, where no other known-ordering in the
literature may satisfy the $G_{1}=0=G_{2}$ condition). In this case, the
position-dependent-particle at hand (2) remains free and admits a
free-particle solution, therefore. However, the triviality of such a choice
(i.e., $G_{1}=0=G_{2}$) inspires the search for yet another feasible
solution for (13) where $G_{1}\neq 0$ (i.e., $\beta \neq -1/2$).

If we just recollect that $\alpha +\beta +\gamma =-1$ (i.e., the von Ross
constraint) and impose the continuity conditions at the abrupt
heterojunction boundaries (i.e., simply the ordering ambiguity parameters $%
\alpha $ and $\gamma $ are related through $\alpha =\gamma $, a manifesto
that ensures the continuity of $m\left( x\right) ^{\alpha }\psi \left(
x\right) $ and $m\left( x\right) ^{\alpha +\beta }\left[ \partial _{x}\psi
\left( x\right) \right] $ at the heterojunction boundary) along with the
choice of $\left( 2G_{1}-G_{2}\right) =0$, we would then dismiss the $\delta
\left( x\right) ^{2}$ ambiguity. Under such conditions, a new set of
ordering ambiguity parameters (the only feasibly admissible within the
current methodical proposal, and the yet to be labeled as MM1-ordering,
hereinafter) that casts $\alpha =\gamma =-3/4$ and $\beta =1/2$ is obtained.
As such and within this new set of ambiguity parameters, the effective
potential (13) collapses into a simple form%
\begin{equation}
V_{eff}\left( q\left( x\right) \right) =\frac{U\left( 0\right) }{2}\delta
^{\prime }\left( x\right) \text{ ; \ }U\left( x\right) =\frac{f^{\prime
}\left( \emph{h}\left( x\right) \right) }{f\left( \emph{h}\left( x\right)
\right) ^{2}}.
\end{equation}

We clearly observe that a scattering problem of a quasi-free quantum
particle (i.e., $V\left( x\right) =0$ whereas $V_{eff}\left( q\left(
x\right) \right) \neq 0$) subjected to the derivative of the one-dimensional
Dirac delta interaction (also called the point dipole interaction) is
manifested by (14) of Hamiltonian (7) (hence, a self-scattering effect
obtains in the process). The detailed solution of which can be inferred from
the scattering coefficients of the $V\left( q\right) =-a\delta \left(
q\right) +b\delta ^{\prime }\left( q\right) $ potential of Gadella et al.
[36] by taking $m=1$, $a=0$ and $b=U\left( 0\right) /2$ as proper parametric
mappings into our model. Choosing to skip all the mathematical and/or
physical details, the reflection and transmission coefficients (see Eq.(23)
of [36]) would , respectively, read%
\begin{equation}
R=-\frac{4U\left( 0\right) }{4+U\left( 0\right) ^{2}}
\end{equation}%
and%
\begin{equation}
T=\frac{4-U\left( 0\right) ^{2}}{4+U\left( 0\right) ^{2}}
\end{equation}%
In a straightforward manner it can be easily shown that the condition $%
\left\vert R\right\vert ^{2}+\left\vert T\right\vert ^{2}=1$ is satisfied.
Consequently, a free quantum particle endowed with the PDM-setting of (2)
may very well experience scattering effects. Moreover, it is obvious that
whilst a $U\left( 0\right) =0$ yields (although trivial) a totally
transparent/reflectionless derivative-of-the-Dirac's delta scatterer, a $%
U\left( 0\right) =\pm 2$ yields a totally reflective
derivative-of-the-Dirac's delta scatterer.

In due course, the position-dependent-mass jump of (1) implies%
\begin{equation}
q\left( x\right) =x\,\sqrt{f\left( \emph{h}\left( x\right) \right) }=x\,%
\left[ 1+\mu \emph{h}\left( x\right) \right] ,
\end{equation}%
and%
\begin{equation}
U\left( x\right) =\frac{\mu }{1+\mu \emph{h}\left( x\right) }\Longrightarrow
U\left( 0\right) =\frac{\mu }{1+\mu /2}.
\end{equation}%
It should be noted here that, in a straightforward manner, one may
substitute $U\left( 0\right) $ in (18) to obtain the transmission and
reflection coefficients (15) and (16), respectively, as%
\begin{equation}
R=-\frac{4\mu \left( 1+\mu /2\right) ^{2}}{4\left( 1+\mu /2\right) ^{4}+\mu
^{2}},
\end{equation}%
and%
\begin{equation}
T=\frac{4\left( 1+\mu /2\right) ^{4}-\mu ^{2}}{4\left( 1+\mu /2\right)
^{4}+\mu ^{2}}.
\end{equation}%
Consequently and asymptotically speaking, a Taylor series expansions about $%
\mu \rightarrow 0$ (i.e., $0<\mu <<1$, also very likely experimentally
feasibly applicable) would result in casting the reflection and transmission
intensities, respectively, as $\left\vert R\right\vert ^{2}\approx \mu
^{2}+O\left( \mu ^{3}\right) $ and\ $\left\vert T\right\vert ^{2}\approx
1-\mu ^{2}+O\left( \mu ^{3}\right) $. Within such asymptotic tendencies, it
is obvious that increasing the value of $\mu $ from just above $0$ to $1$
would make the derivative-of-the-Dirac's delta scatterer less transparent
until total reflection takes place when $\mu =1$.

\section{Concluding remarks}

In this work, a quasi-free quantum particle (i.e., $V\left( x\right) =0$
whereas $V_{eff}\left( q\left( x\right) \right) \neq 0$)\ endowed with a
position-dependent-mass jump (1) is considered. Whilst, under Mustafa's and
Mazharimousavi's ordering ambiguity parametrization (i.e., $\alpha =\gamma
=-1/4$ and $\beta =-1/2$) the free PDM quantum particle remained free, we
have witnessed (only under a specific though rather new set of ordering
ambiguity parametrization $\alpha =\gamma =-3/4$ and $\beta =1/2$) that
scattering effects are manifested by the particle's by-product introduction
of the derivative-of-Dirac's delta function, $\delta ^{\prime }\left(
x\right) $, in the PCT process. We were able to obtain the related
reflection (15) and (16) transmission coefficients for any $U\left( 0\right) 
$ of $U\left( x\right) $ in (14). We have predicted that a quasi-free PDM
quantum particle may very well totally reflect itself (documented in
(17)-(20)) by the effective potential it introduces (i.e., the
derivative-of-Dirac's delta function, $\delta ^{\prime }\left( x\right) $,
in this case).

Unavoidably, nevertheless, it should be noted that if the PDM quantum
particle at hand (1) is also subjected to Dirac delta potential $V\left(
x\right) =-a\delta \left( x\right) $ that appears in equation (1) of Gadella
et al. [36] (though rather readily laid far beyond our methodical proposal
above), then the reflection and transmission coefficients would,
respectively, read%
\begin{equation}
R=-\frac{4\left[ a+ikU\left( 0\right) \right] }{4a+ik\left[ 4+U\left(
0\right) ^{2}\right] },
\end{equation}%
and%
\begin{equation}
T=\frac{ik\left[ 4-U\left( 0\right) ^{2}\right] }{4a+ik\left[ 4+U\left(
0\right) ^{2}\right] }.
\end{equation}%
The comprehensive discussion of which is given by Gadella et al. [36].

Finally, we have very recently shown (see Mustafa and mazharimousavi [39]
for more details) that the ordering ambiguity conflict (associated with the
non-unique representation of the von Roos PDM Schr\"{o}dinger Hamiltonian)
as to which ordering would be the best representative for the PDM
Hamiltonian, can not be resolved within the abrupt heterojunction boundary
conditions and the Dutra's and Almeida's [9] reliability test. The PDM forms
have their say in the process. The current Heaviside-dependent mass
form/jump (2) was not just an amazing example but yet an additional
documentation of this observation.

\end{document}